\documentclass[aps,pra,twocolumn]{revtex4}
 \usepackage[usenames,dvipsnames]{color}
\usepackage{graphicx}
\usepackage{dcolumn}
\usepackage{bm}
\usepackage{amsmath}
\usepackage{amsfonts}
\usepackage{psfrag}

\def\bra#1{\mathinner{\langle{#1}|}} 
\def\ket#1{\mathinner{|{#1}\rangle}}

\newcommand{\Eq}[1]{Eq.~(\ref{#1})}

\begin{document}
\title{Perspectives for gapped bilayer graphene polaritonics}

\author{Simone De Liberato}
\affiliation{School of Physics and Astronomy, University of Southampton, Southampton SO17 1BJ, United Kingdom}

\begin{abstract}
Bilayer graphene is normally a semimetal with parabolic dispersion, but a tunable bandgap up to few hundreds meV can be opened by breaking the symmetry between the layers through an external potential. Ab-initio calculations show that the optical response around the bandgap is strongly dominated by bound excitons, whose characteristics and selection rules differ from the usual excitons found in semiconductor quantum wells. 
In this work we study the physics of those excitons resonantly coupled to a photonic microcavity, assessing the possibility to reach the strong and the ultrastrong coupling regimes of light-matter interaction. We discover that both regimes are experimentally accessible, thus opening the way for a most promising technological platform, combining mid-infrared quantum polaritonics with the tunability and electronic features of graphene bilayers.
\end{abstract}
\maketitle

\psfrag{LE}[cr][tr][0.8]{(a)}
\psfrag{LL}[cr][tr][0.8]{(b)}

\psfrag{LO}[cr][tr][0.8]{(a)}
\psfrag{LC}[cr][tr][0.8]{(b)}

\psfrag{XE}[tc][bc][0.8]{$V_{\text{ext}}$ (meV)}
\psfrag{XL}[tc][bc][0.8]{$V_{\text{ext}}$ (meV)}

\psfrag{YE}[Bc][tc][0.8]{$\hbar\omega_x$ (meV)}
\psfrag{YL}[Bc][tc][0.8]{Absorption}

\psfrag{TE}[cl][cc][0.8]{\hspace{-0.6cm} $\hbar\omega_x=c_2V_{\text{ext}}^2+c_1V_{\text{ext}}$}
\psfrag{TL}[cl][cc][0.8]{$\nu=dV_{\text{ext}}$}

\psfrag{XO}[tc][bc][0.8]{$V_{\text{ext}}$ (meV)}
\psfrag{XC}[tc][bc][0.8]{$L_{\text{eff}}$ (in units of $\lambda$)}

\psfrag{YO}[Bc][tc][0.8]{$\Omega_R/\omega_x$ ($\times 10^{-2}$)}
\psfrag{YC}[Bc][tc][0.8]{$\Omega_R/\omega_x$}

\psfrag{TC}[cc][cc][0.8]{$V_{\text{ext}}=400$ meV}
\psfrag{TO}[cc][cc][0.8]{$L_{\text{eff}}=\frac{\lambda}{2}$}

\section{Introduction}

Microcavity polaritons are half-light and half-matter quasiparticles created by the strong coupling between a confined photonic mode and a matter excitation \cite{Kavokin,Carusotto13}. Thanks to their hybrid nature they have many unique properties, making them the object of intense research for applications as different as low threshold lasers \cite{Schneider13,Bhattacharya14}, terahertz emitters \cite{Kavokin12,DeLiberato13,Liew13,Barachati15}, and quantum simulators \cite{Byrnes10,Jacqmin14,Sala15}.
Since their initial discovery in GaAs quantum wells \cite{Weisbuch92}, various kinds of polaritonic resonances have been observed in many solid-state cavity quantum electrodynamics setups, ranging from organic microcavities \cite{Lidzey98}, to intersubband transitions in doped quantum wells \cite{Dini03}, or more recently Landau level systems \cite{Scalari12}, and transition-metal dichalcogenides \cite{Liu15}.

While microcavity polaritons are not observable in intrinsic monolayer graphene, due to the absence of discrete excitonic resonances, there have been multiple proposals to obtain polaritonic effects in graphene-based systems, either by gapping a single graphene layer \cite{Berman12}, applying strong magnetic fields \cite{Hagenmuller12}, or coupling two different graphene sheets separated by a dielectric slab \cite{Zhang08,Lozovik08,Bistritzer08,Bistritzer08b,Kharitonov08,Berman12b}.  While none of these proposals has been implemented so far, there is little doubt that the observation of graphene-based microcavity polaritons  would be an important milestone for present-day nanophotonics. Merging two very powerful and versatile technological platforms as graphene bilayers and microcavity polaritons could lead to major breakthroughs for mid-infrared optoelectronics and quantum technologies. In particular strong coupling, by increasing the amount of light that graphene can absorb or emit, could overcome the main obstacle that limits the exploitation of graphene in optoelectronic devices \cite{Gerstner12, Engel12,Furchi12}.

\begin{figure}[h!]
\begin{center}
\hspace{-0.73cm}
\includegraphics[width=9.25cm]{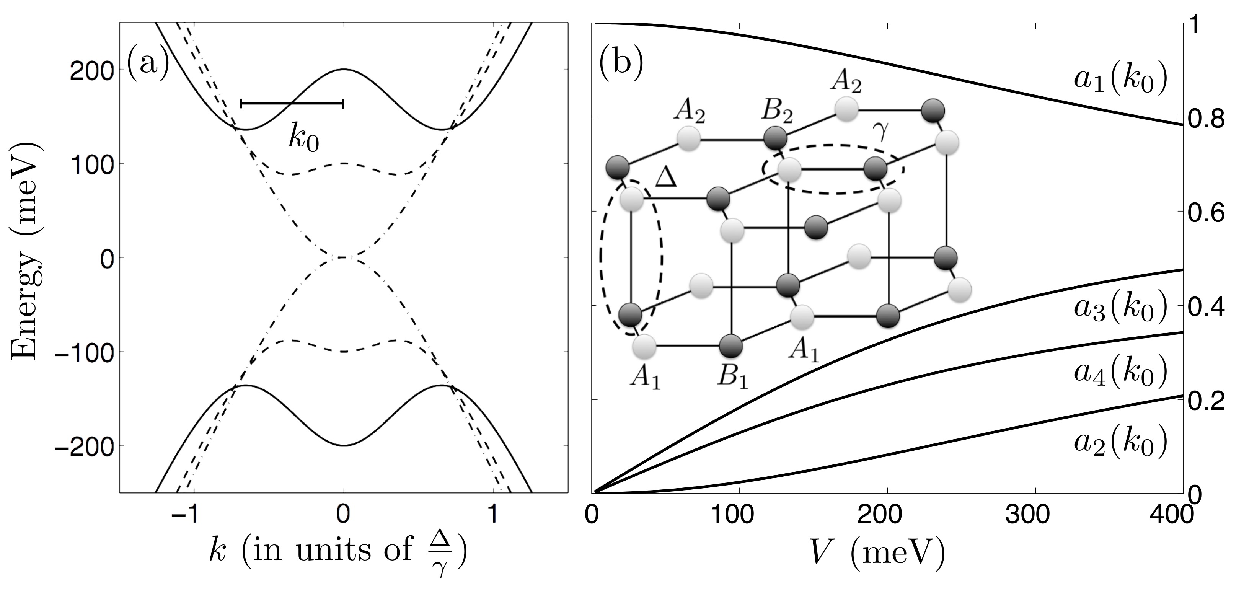} 
\caption{\label{Fig1} Left panel: dispersion of the two lowest energy bands in graphene bilayer without asymmetry (dash-dotted line), an asymmetry parameter $V=200$ meV (dashed line) and $V=400$ meV (solid line). For this last line the Mexican-hat shape of the dispersion is evident and the $k_0$ point is explicitly shown.
Right panel: coefficients of the wavefunction as a function of the asymmetry, as from \Eq{wavevector}. In the inset we plot instead a scheme of the bilayer with its tight-binding parameters.}
\end{center}
\end{figure}

In this paper we will investigate the possibility to observe polaritonic effects in a different graphene-based system, that is a bilayer composed of two graphene layers in the AB (Bernal) stacking (see the inset in Fig. \ref{Fig1}(b)), the kind of bilayer obtained by standard graphite exfoliation. As intrinsic graphene, bilayer graphene is normally a semimetal, but the interaction between the two layers results in parabolic bands around the $K$ and $K'$ points. When the symmetry between the two layers is broken, e.g., through an externally applied potential, a tunable bandgap, up to around $300$ meV, can be opened  \cite{McCann06,McCann06b,Ohta06,Oostinga07,Ando07,Ando09,CastroNeto09,Yang10,McCann13,Zhang09,Mak09,Droscher12}. 
In Fig. \ref{Fig1}(a) we plot the dispersion of the first valence and conduction bands for different values of the asymmetry potential $V$. 
We see that, when a gap is present, the shape of the dispersion is not parabolic anymore, but it assumes a typical Mexican-hat shape, with the band edge on a circle of radius $k_0$ in momentum space.  
Ab-initio calculations show that the optical response of gapped bilayer graphene (GBG) around the bandgap is strongly dominated by bound excitons \cite{Park10,Liang12}.
Those excitons are of the Wannier kind, delocalised over many carbon atoms on both layers, and they present various peculiarities that distinguish them from excitons in other two dimensional materials. In particular, due to its peculiar dispersion, GBG presents a van Hove singularity, typical of one dimensional systems, in the joint electronic density of states around the badgap. Not only this diverging density of states leads to rather large excitonic features, but the $n_w=2$ pseudospin winding number of GBG \cite{Park11} also strongly modifies optical selection rules \cite{Ando07,Ando09}.

In the following we will study the physics of those excitons resonantly coupled with the photonic mode of a microcavity. Our principal aim will be to estimate the maximal achievable ratio between the strength of the light-matter interaction, quantified by the vacuum Rabi frequency, $\Omega_R$, and the bare frequency of the excitonic resonance, $\omega_x$. Such a quantity, that we will call normalised coupling, will give us two fundamental pieces of information, allowing us to assess the scientific and technological interest of this novel solid-state cavity quantum electrodynamics setup.
 The first  information is the minimal quality factor $Q$ of the photonic resonator which allows the system to be in the strong light-matter coupling regime, and thus to sustain polaritonic resonances. In order to be in the strong coupling regime the
vacuum Rabi frequency needs to be larger than the losses, so that it becomes possible to spectroscopically resolve the resonant splitting between the two coupled modes \cite{Kavokin}. Given the large excitonic binding energies of excitons in GBG \cite{Park10} it is safe to assume that the leading losses are due to the photonic part of the polaritonic field and, remembering that the quality factor of a resonator is the ratio between its frequency and its loss rate, the condition to be in the strong coupling regime thus reads 
\begin{eqnarray}
\label{Q}
\tfrac{\Omega_R}{\omega_x}>\tfrac{1}{4Q}.
\end{eqnarray}
The second information is the extent to which higher order photonic processes are observable in GBG polaritons. Applying standard perturbation theory to calculate the effects of light-matter coupling we find in fact that the $n^{\text{th}}$ order perturbative term is proportional to the interaction to the power of $n$, divided the energy gap to the power of $n-1$, that is it will scale as $\omega_x \left(\tfrac{\Omega_R}{\omega_x} \right)^{n}$.
The normalised coupling $\tfrac{\Omega_R}{\omega_x}$ is thus the relevant dimensionless parameter quantifying the coupling between light and matter. 
If such a dimensionless parameter becomes non-negligeable, a regime referred to as ultrastrong coupling regime \cite{Ciuti05,Devoret07}, many novel physical phenomena due to higher order processes become observable, ranging from quantum vacuum radiation \cite{DeLiberato07} and quantum phase transitions \cite{Lambert04,Nataf10b}, to the modification of energy transport \cite{Orgiu15,Feist15} and light emission \cite{DeLiberato14,Ripoll15,Bamba15} properties, to the appearance of cavity assisted chemical and thermodynamical effects \cite{Hutchison12,Hutchison13,Galego15,Cwik15}. 
Notice that the transition between strong and ultrastrong coupling regimes is a smooth crossover, determined more by experimental sensitivity to those novel phenomena that by any intrinsic qualitative change in the underlying physics. The boundary between the two, that is conventionally fixed at  $\tfrac{\Omega_R}{\omega_x}=0.1$, was in fact the coupling value that allowed for the first observation of ultrastrong coupling effects \cite{Anappara09}.
The ultrastrong coupling regime has been since then observed in a variety of different systems \cite{Niemczyk10,Muravev11,Geiser12,Schwartz11,Porer12,Askenazi14,Gambino14,Gubbin14}, with an actual record of $\tfrac{\Omega_R}{\omega_x}= 0.87$ \cite{Maissen14}. 
Excitons in GBG could be very interesting in this regards, a priori allowing to reach or even improve such a record. On one hand thanks to their  peculiar optical properties, 
leading to very strong absorption peaks \cite{Park10}, on the other hand thanks to the nanometric dimension of the carbon bilayer, that can thus be placed very close to the metallic surface of metamaterial resonators, achieving very large field enhancements \cite{Maier06,Benz13}.
The rest of this paper is structured as follows. In Sec. \ref{Theory} we will lay down the theory of the coupling between excitons in GBG and the photonic mode of a resonator. In Sec. \ref{Numerical} we will use the numerical results of Ref. \cite{Park10} to estimate the parameters we need in order to fully characterise the system. With a solid theory and reliable values for the parameters, in Sec. \ref{Results} we will then calculate the normalised light-matter coupling, showing how both the strong and the ultrastrong coupling regimes are experimentally accessible using GBG. Finally, in Sec. \ref{Conclusions}, we will draw some conclusions, also pointing out directions for future investigations.

\section{Theory of light-matter coupling for GBG excitons}
\label{Theory}
Graphene is an hexagonal lattice of carbon atoms, having
two atoms per unit cell, usually labelled $A$ and $B$. Around the two inequivalent $K$ and $K'$ points at the corners of the Brillouin zone it presents a linear dispersion, from which most of its exotic electronic properties stem \cite{CastroNeto09}.
A graphene bilayer in AB stacking is obtained by putting two graphene monolayers one over the other, such that the $A$ atom of one layer is on the top of the $B$ atom of the other.
In a tight-binding approach the wavefunction of a graphene bilayer can be written as a four-component vector, describing the wavefunction amplitude in each of the two inequivalent sublattices $\{A,B\}$ in each layer $\{1,2\}$ \cite{Ando09,McCann13}. Around the $K$ point the Hamiltonian can be written in the $(A_1, B_1, A_2, B_2)$ basis, keeping only the dominant next-neighbour terms, as the direct sum of the Hamiltonian for each graphene monolayer (with hopping parameter $\gamma$) plus a term proportional to $\Delta$, describing the interlayer coupling between the dimer sites $B_1$ and $A_2$
\begin{eqnarray}
\label{H0K}
H^K_0&=&
\left(
\begin{array}{cc}
 \tfrac{V}{2}+\gamma\boldsymbol{\sigma}\cdot\mathbf{p} &\Delta\sigma^-      \\
 \Delta\sigma^+&-\tfrac{V}{2}+\gamma\boldsymbol{\sigma}\cdot\mathbf{p}   
\end{array}
\right),
\end{eqnarray}
where $V$ is a parameter describing the asymmetry between the two layers, $\boldsymbol{\sigma}=\sigma_x \boldsymbol{\epsilon}_x+\sigma_y \boldsymbol{\epsilon}_y$, with $\sigma_{x,y}$ the Pauli matrices, $\sigma^{\pm}=\tfrac{1}{2}(\sigma_x\pm i\sigma_y)$,
and $\boldsymbol{\epsilon}_{x,y}$ are linear polarisation versors. Such an Hamiltonian can be analytically diagonalised, leading to the following dispersion for the lowest valence and conduction bands as a function of the in-plane wavevector $\mathbf{k}$ \cite{Ando09}
\begin{eqnarray}
\label{Evc}
E_{v,c}^2&=&\frac{V^2+2\Delta^2}{4}+\gamma^2k^2-\sqrt{(V^2+ \Delta^2)\gamma^2 k^2+\tfrac{\Delta^4}{4}},\quad
\end{eqnarray}
whose shape is plotted in Fig. \ref{Fig1}(a) for various values of $V$, and whose eigenvectors can be written as
\begin{eqnarray}
\label{wavevector}
\mathbf{F}_{c,\mathbf{k}}(\boldsymbol{\rho})&=&\frac{e^{i\mathbf{k\cdot \boldsymbol{\rho}}}}{\sqrt{S}} 
\lbrack a_1(k),a_2(k) e^{i\theta_{{k}}},a_3(k) e^{i\theta_{{k}}},a_4(k) e^{2i\theta_{{k}}}\rbrack,\\
\mathbf{F}_{v,\mathbf{k}}(\boldsymbol{\rho})&=&\frac{e^{i\mathbf{k\cdot \boldsymbol{\rho}}}}{\sqrt{S}} 
\lbrack -a_4(k),a_3(k) e^{i\theta_{{k}}},-a_2(k) e^{i\theta_{{k}}},a_1(k) e^{2i\theta_{{k}}}\rbrack,\nonumber
\end{eqnarray}
with $\boldsymbol{\rho}$ the in-plane position vector and $S$ the sample surface. 
In Fig. \ref{Fig1}(a) we clearly see the Mexican-hat dispersion, with the band edge at a finite value of the in-plane wavevector, that can be calculated by differentiating  \Eq{Evc} as
\begin{eqnarray}
k_0^2=\tfrac{V^4+2V^2\Delta^2}{4(V^2+\Delta^2)}.
\end{eqnarray}
We expect that the low energy excitons will mainly consist of linear superpositions of electronic transitions localised around the band edge \cite{Skinner14}, an hypothesys at least qualitatively confirmed by the numerical data of Ref. \cite{Liang12}. In Fig. \ref{Fig1}(b) we thus plot the $a_j(k)$ coefficients at $k=k_0$, $j\in\{1,2,3,4\}$. We see that the $a_1(k_0)$ coefficient is the dominant one, at least for moderate values of the asymmetry. In the following, in order to simplify the theory, we will thus always assume
\begin{eqnarray}
\label{approx}
a_1(k)&\gg& a_2(k),a_3(k),a_4(k).
\end{eqnarray}
This approximation implies that valence and conduction electrons are mainly localised on different layers, in the non-dimer sites $A_1$ and $B_2$, a conclusion supported by the numerical results of Refs. \cite{Park10,Liang12}. 

The interaction Hamiltonian can be derived by applying the usual minimal-coupling substitution $\mathbf{p}\rightarrow \mathbf{p}+e\mathbf{A}(\mathbf{r})$ in \Eq{H0K}, where $\mathbf{A}(\mathbf{r})$ is the electromagnetic vector potential and $\mathbf{r}=\{\boldsymbol{\rho},z\}$ is the three dimensional position vector.  We thus obtain
\begin{eqnarray}
\label{Hint}
H_{\text{int}}&=&\frac{e\gamma}{\hbar}\mathbf{A}(\mathbf{r})\cdot
\left(
\begin{array}{cc}
    \boldsymbol{\sigma}  &0    \\
 0& \boldsymbol{\sigma} 
\end{array}
\right).
\end{eqnarray}
Note that in \Eq{Hint} there is no dipole along $z$, that is consisten with our approximation of considering the electrons localised around the non-dimer states, thus not allowing for currents normal to the bilayer plane.
We will thus limit ourselves to the case of normal incidence, for which the coupling is maximised as the field lies completely in the dipole plane.
Introducing the second quantised annihilation operators for electrons, $e_{\mathbf{k}}$, and holes, $h_{\mathbf{k}}$, with in-plane wavevector $\mathbf{k}$, the exciton state with zero center of mass momentum around the $K$ point can be written as 
\begin{eqnarray}
\ket{\psi}&=&\sum_{\mathbf{k}} \psi(\mathbf{k}) e^{\dagger}_{\mathbf{k}}h^{\dagger}_{-\mathbf{k}}\ket{0},
\end{eqnarray}
with $\ket{0}$ the crystalline ground state.
We can now calculate the interaction matrix element in the dipole approximation as
\begin{eqnarray}
\label{Melement}
\bra{0}H_{\text{int}}\ket{\psi}&=&\frac{e\gamma}{\hbar}\boldsymbol{A}\cdot
\sum_{\mathbf{k}} \psi(\mathbf{k})\bra{0}\left(
\begin{array}{cc}
    \boldsymbol{\sigma}  &0    \\
 0& \boldsymbol{\sigma} 
\end{array}
\right) e^{\dagger}_{\mathbf{k}}h^{\dagger}_{-\mathbf{k}}\ket{0}
\nonumber \\&=&\frac{\sqrt{2^3}e\gamma}{\hbar}\boldsymbol{A}\cdot
\sum_{\mathbf{k}} \psi(\mathbf{k})\lbrack \boldsymbol{\epsilon}_+ a_1(k)a_3(k) e^{-i\theta_{k}}\nonumber \\&&
- \boldsymbol{\epsilon}_- a_2(k)a_4(k) e^{i\theta_{k}} \rbrack,
\end{eqnarray}
with 
\begin{eqnarray}
\boldsymbol{\epsilon}_{\pm}&=&\frac{\boldsymbol{\epsilon}_x\pm i \boldsymbol{\epsilon}_y}{\sqrt{2}},
\end{eqnarray}
the circular polarisation versors and
\begin{eqnarray}
\boldsymbol{A}&=&\frac{1}{S}\int d\boldsymbol{\rho} \,\mathbf{A}(\boldsymbol{\rho},z_0),
\end{eqnarray}
where $z_0$ is the position of the GBG plane.
As firstly noted in Ref. \cite{Park10}, the phases of the electron and hole wavefunctions  in \Eq{wavevector}, leading to a pseudospin winding number $n_w=2$ \cite{Park11}, modify selection rules for excitons. The sum over the $\theta_k$ in 
 $\sum_{\mathbf{k}} \psi(\mathbf{k})e^{\pm i\theta_{k}}$ implies that only excitons with angular dependency proportional to $e^{\pm i\theta_{k}}$, that is having envelope angular momentum $m_{\text{env}}=\pm1$, are coupled to the crystalline ground state.
Under the hypothesis expressed in \Eq{approx}, we can thus consider that effectively only the $m_{\text{env}}=1$ exciton couples around the $K$ point. Considering the four-fold spin and valley multiplicity, and taking care that for excitons around the $K'$ point we need to consider the inversion $y\rightarrow -y$, leading to have the $m_{\text{env}}=-1$ exciton coupled instead \cite{Park10}, we thus obtain the following interaction Hamiltonian
\begin{eqnarray}
\label{Hint2}
H_{\text{int}}&=&\frac{4e\gamma\nu\sqrt{S}}{\hbar}\boldsymbol{A}\cdot
\sum_{\sigma}
\boldsymbol{\epsilon}_{\sigma}(b^{\dagger}_{\sigma}+b_{\sigma}),
\end{eqnarray}
with
\begin{eqnarray}
\nu&=&\frac{1}{\sqrt{S}}\sum_{\mathbf{k}} \psi(\mathbf{k})e^{-i\theta_{k}} a_1(k)a_3(k),
\end{eqnarray}
a parameter depending on the band structure and on the exciton wavefunction, and $b_{\sigma}$ the annihilation operator for an exciton of polarisation $\sigma$.

For the photonic cavity we will use a general model, taking into account the possibility to use subwavelength confinement in order to increase the coupling. We can write the electromagnetic cavity field for a single mode of frequency $\omega_c$ as
\begin{eqnarray}
\boldsymbol{A}&=&\sqrt{\frac{\hbar}{\epsilon_0\omega_c SL_{\text{eff}}}}(a+a^{\dagger})\boldsymbol{\epsilon},
\end{eqnarray}
where $a$ is the annihilation operator for a photon, $\boldsymbol{\epsilon}$ its polarisation at the location of the GBG,
and  the effective length, $L_{\text{eff}}$, is defined as \cite{Maier06}
\begin{eqnarray}
L_{\text{eff}}\,u(z_0)^2=\int dz\, \epsilon_r(z)\,u(z)^2,
\end{eqnarray}
where $\mathbf{u}(z)$ is the profile of the electromagentic mode.
The effective length quantifies to which extent the energy of a cavity photon is
concentrated around the graphene bilayer, and it is thus a crucial element to
determine the strength of the light-matter interaction: the shorter the effective length,
the higher the field at the location of the excitons, the larger the coupling.
Considering only the excitonic mode that couples to the photonic one we thus finally get
\begin{eqnarray}
H_{\text{int}}&=&
\frac{4e\gamma\nu}{\sqrt{\epsilon_0\hbar\omega_c L_{\text{eff}}}}(a+a^{\dagger})(b^{\dagger}+b),
\end{eqnarray}
from which we can directly read the normalised coupling at resonance
\begin{eqnarray}
\label{normcoupling}
\frac{\Omega_R}{\omega_x}=
\frac{4e\gamma\nu}{\sqrt{\epsilon_0(\hbar\omega_x)^3 L_{\text{eff}}}}.
\end{eqnarray}
\begin{figure}
\begin{center}
\hspace{-0.4cm}
\includegraphics[width=4.6cm]{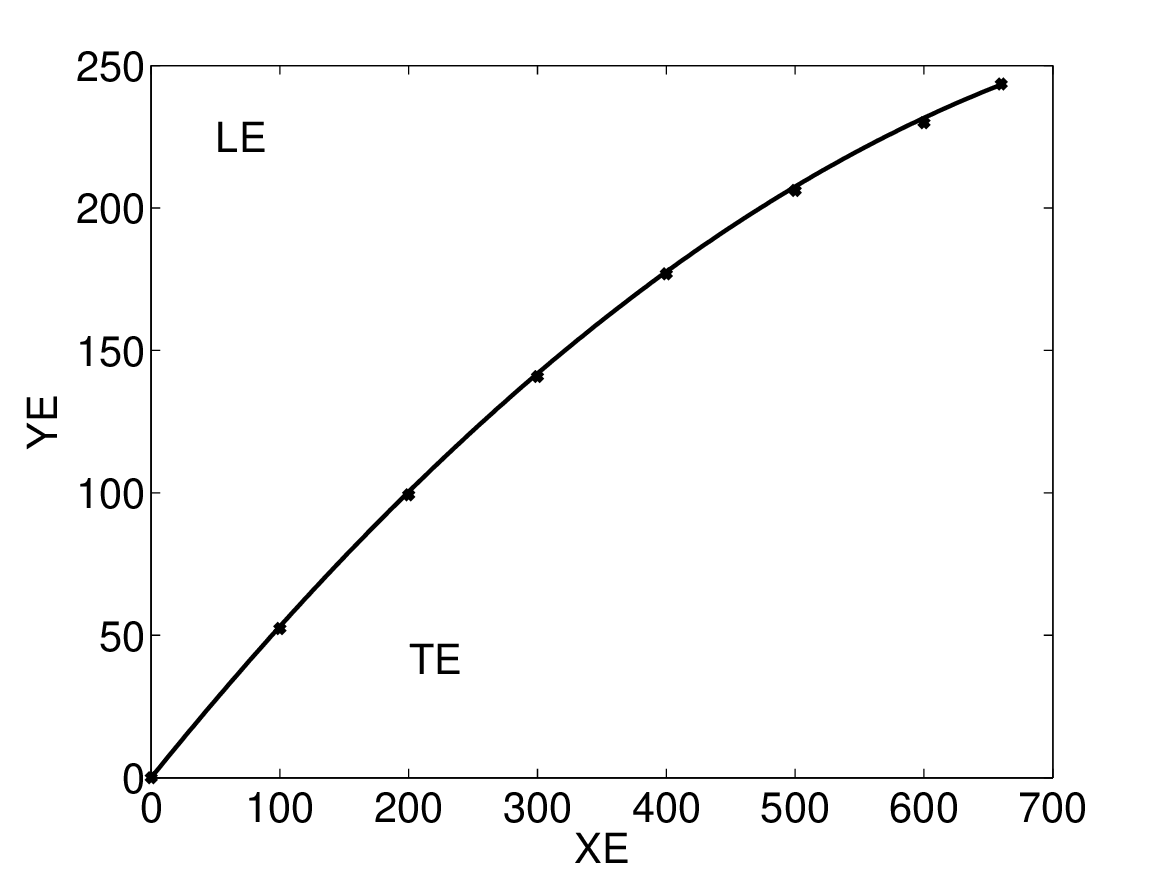} 
\hspace{-0.5cm}
\includegraphics[width=4.6cm]{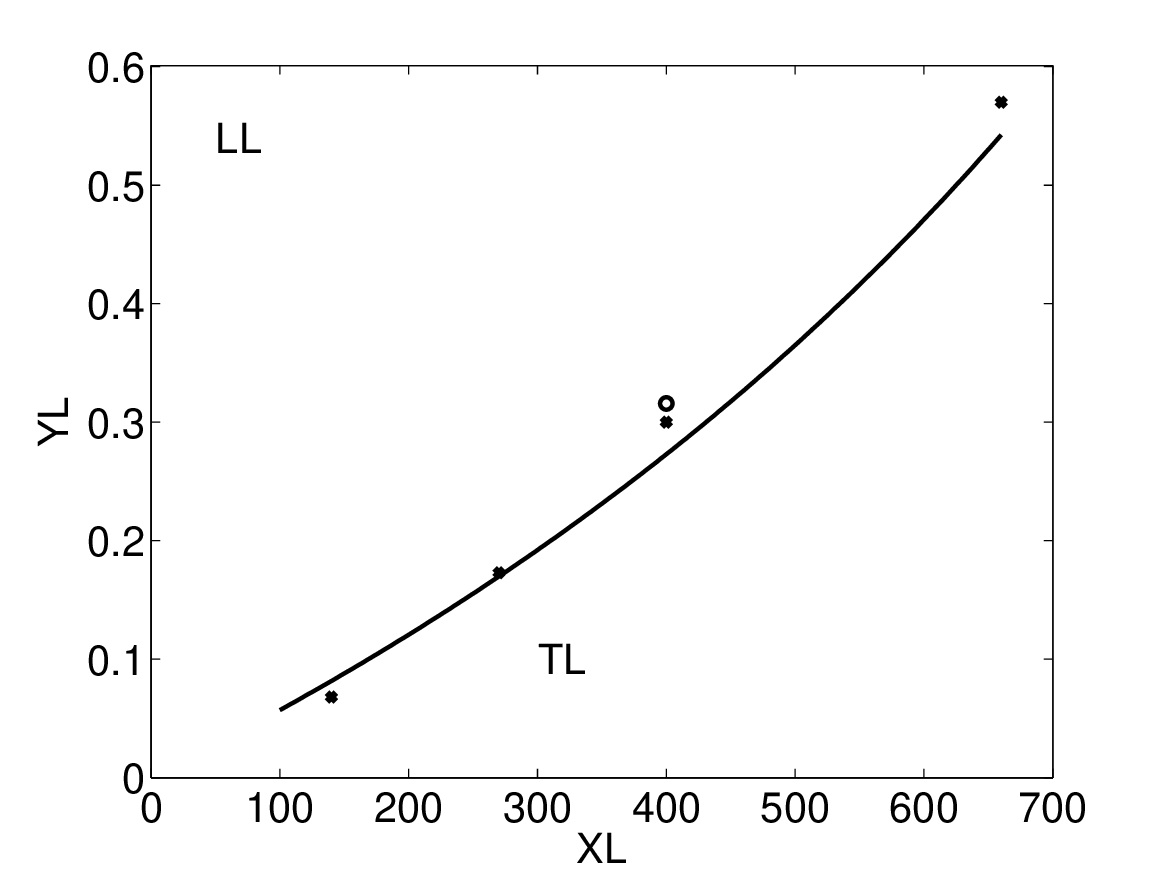} 
\caption{\label{Fig2} Left panel: fit of the exciton energy as a function
of the applied potential $V_{\text{ext}}$. The crosses are
numerical data taken from the Fig. 4(a) of Ref. \cite{Park10} and the continuous line the numerical fit done using \Eq{Efit}. 
Right panel: fit of the maximal absorption of the excitonic peak as a function of the applied potential $V_{\text{ext}}$.
The crosses are numerical data taken from the Fig. 3 of Ref. \cite{Park10}, the continuous line the numerical fit done using \Eq{Efit} and \Eq{AbsP},
and the circle is the point derived from the integrated absorption in 
 \Eq{AbsI}.}
\end{center}
\end{figure}

\section{Parameters estimation}
\label{Numerical}

In order to extract useful numerical values from \Eq{normcoupling}, able to tell us if either the strong or the ultrastrong coupling regimes are observable with GBG, we need to have reliable estimates for the coupling parameter $\nu$, and for the excitonic energy $\hbar\omega_x$. While it is a priori possible to solve the Bethe-Salpeter equation following, for example, an approximation scheme similar to the one used in Ref. \cite{Skinner14}, the results obtained in this way are not in quantitative agreement with the ab-initio predictions from Refs. \cite{Park10,Liang12}. This does not come as a surprise given that the method developed in Ref. \cite{Skinner14} assumes vanishingly small binding energies, while the binding energies obtained from ab-initio calculations are sizeable fractions of the bandgap.

In order to get reliable estimates of the achievable normalised couplings, we will thus use data from Ref. \cite{Park10} to obtain fits for $\hbar\omega_x$ and $\nu$. Notice that in such a reference the results are all presented not as functions of the asymmetry parameter $V$ that appears in the Hamiltonian in \Eq{H0K}, but of the physically applied external potential $V_{\text{ext}}$. While one can a priori be calculated in function of the other, as shown in Ref. \cite{McCann06}, resulting in a $V$ of the order of $V_{\text{ext}}/2$, we do not need to perform such a conversion here, because all the parameters we need are already given as functions of $V_{\text{ext}}$. For consistency we take also the band structure parameters from the same reference, even if a few different values are reported in the literature \cite{McCann13}, that is we will consider $\gamma=5.4$ eV\AA $\,$
and $\Delta=0.37$ eV.

The exciton energy $\hbar\omega_x$ (the optical bandgap) is plotted in Fig. 4(a) of Ref. \cite{Park10} as a function of $V_{\text{ext}}$. We fitted such a plot using a quadratic function 
\begin{eqnarray}
\label{Efit}
\hbar \omega_x&=&c_2V_{\text{ext}}^2+c_1V_{\text{ext}},
\end{eqnarray}
that, as shown in Fig. \ref{Fig2}(a), gives a very good fit for $c_2=-2.9\times 10^{-4}$ meV$^{-1}$ and $c_1=0.56$.
To estimate $\nu$ instead we can exploit the Fig. 3 of Ref. \cite{Park10}, where the excitonic absorption for different values of $V_{\text{ext}}$ is plotted assuming an arbitrary HWHM broadening $\Gamma=5$ meV. From \Eq{Hint2} we can calculate the absorbed power for a normally incident plane wave of frequency $\omega$ and vector potential $\mathbf{A(r)}$ using the Fermi golden rule 
\begin{eqnarray}
\label{AbsP}
P_{\text{abs}}&=&\frac{32\pi e^2 \gamma^2\mathbf{A(r)}^2\nu^2 S }{\hbar^2}\rho(\omega),
\end{eqnarray}
with $\rho(\omega)$ the density of states of the exciton state, a Lorentzian of width $\Gamma$. Dividing \Eq{AbsP} by the incident flux
\begin{eqnarray}
P_{\text{in}}&=&\frac{\epsilon_0 c \hbar \omega}{2}\mathbf{A(r)}^2S,
\end{eqnarray}
we obtain the absorption spectrum
\begin{eqnarray}
\label{Abs}
A(\omega)&=&\frac{64\pi e^2 \gamma^2 \nu^2}{\epsilon_0 c \hbar^3 \omega}  \rho(\omega)=\frac{256\pi^2\alpha \gamma^2 \nu^2 }{\hbar^2\omega} \rho(\omega),
\end{eqnarray}
where $\alpha$ is the fine structure constant. At resonance $\rho(\omega)=\frac{1}{\pi\Gamma}$ and, knowing the excitonic frequencies from \Eq{Efit}, we can thus fit $\nu$ against the maximal absorption for various values of $V_{\text{ext}}$. In Fig. \ref{Fig2}(b) we show that a good fit can be obtained with a linear approximation
\begin{eqnarray}
\label{nufit}
\nu&=&d V_{\text{ext}},
\end{eqnarray}
with $d=2.9\times 10^4$ meV$^{-1}$m$^{-1}$.
The consistency of such a fit can be tested against the value of the integrated absorption for the lowest bright exciton, IA$=1.24$ meV, for $V_{\text{ext}}=400$ meV, explicitly given in Ref. \cite{Park10}. Integrating \Eq{Abs} we obtain
\begin{eqnarray}
\label{AbsI}
4\times\text{IA}&=&\frac{64\pi^2\alpha \gamma^2 \nu^2 }{\hbar^2\omega},
\end{eqnarray}
where the factor $4$ is due to the four-fold exciton degeneracy.
The obtained value is shown as a circle in Fig. \ref{Fig2}(b).

\section{Numerical results}
\label{Results}
\begin{figure}
\begin{center}
\hspace{-0.5cm}
\includegraphics[width=4.6cm]{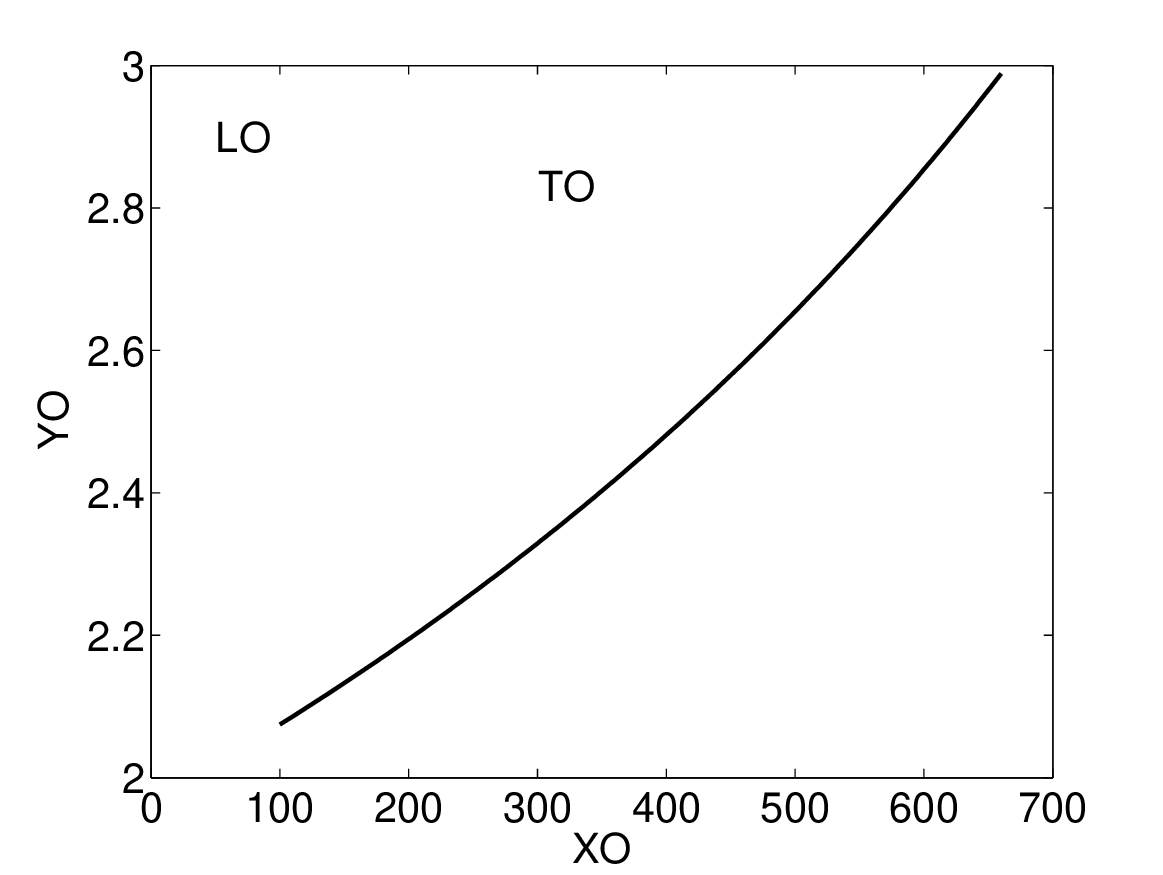} 
\hspace{-0.4cm}
\includegraphics[width=4.6cm]{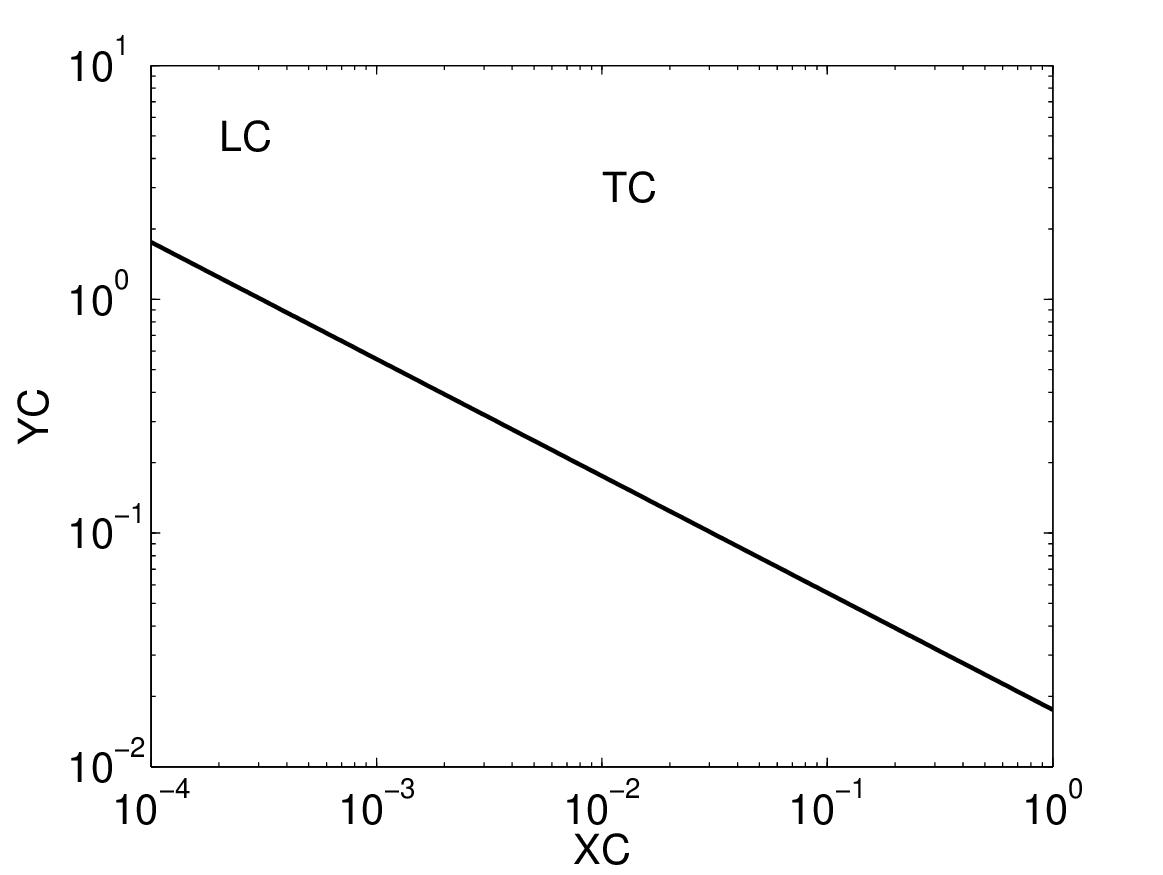} 
\caption{\label{Fig3} Left panel: normalised coupling as a function of the applied potential $V_{\text{ext}}$, for a resonant $\tfrac{\lambda}{2}$ cavity. Right panel: normalised coupling as a function of the effective cavity length $L_{\text{eff}}$, for $V_{\text{ext}}=400$ meV. }
\end{center}
\end{figure}
Plugging \Eq{Efit} and \Eq{nufit} into \Eq{normcoupling} we can finally
calculate the normalised coupling as a function of the external potential. 
In Fig. \ref{Fig3}(a) we plot the normalised coupling as a function $V_{\text{ext}}$ for a $\tfrac{\lambda}{2}$ cavity, that is choosing $L_{\text{eff}}=\tfrac{c \pi}{\omega_x}$. We see that normalised couplings of the order or $\tfrac{\Omega_R}{\omega_x}\simeq 0.03$ are achievable, that from \Eq{Q} implies that quality factors as low as $Q=10$ would allow the system to be in the strong coupling regime. This is a very bland requirement given that, in the mid-infrared, quality factors $Q>10^3$ have been achieved using dielectric cavities based on narrow bandgap IV-VI semiconductors \cite{Heiss01}. Moreover, graphene layers have already successfully been coupled to both metallic \cite{Engel12} and dielectric \cite{Furchi12} $\frac{\lambda}{2}$ planar cavities. While those experiments were performed at shorter wavelengths, they are nevertheless important to assess the feasibility of such a kind of device.
In Fig. \ref{Fig3}(b) we plot instead the normalised coupling as a function of the effective length $L_{\text{eff}}$, normalised to the free space wavelength $\lambda$. We see that the ultrastrong coupling regime is achievable already for subwavelength confinements $L_{\text{eff}}\simeq 0.03 \lambda$, and that very high values are achievable for tighter ones.
From Ref. \cite{Maier06}, where a simple condenser-type cavity is studied, we see that $L_{\text{eff}}$ small enough are achievable while still having a sizeable quality factor. Notice that linear subwavelength confinements, of the order of $10^{-2}$ in the mid-infrared, corresponding to $\tfrac{\Omega_R}{\omega_x}\simeq 0.2$, are commonly achieved in metallic structures \cite{Benz13} or, more recently, in localised surface phonon polariton samples \cite{Caldwell13,Chen14}, where such low mode volumes are accompanied by quality factors $Q>100$, over an order of magnitude larger of what necessary to resolve the polaritonic resonances. Mid-infrared subwavelength resonators have also already been coupled to single graphene layers \cite{Yao14}.
Notice that, contrary to other kinds of mid-infrared polaritonic structures based on intersubband transitions, where the effective field confinement is limited by the fact that it is not possible to put the two dimensional electron gas closer than a few tens of nanometers from the metallic plasmonic surface \cite{Benz13}, in our case the GBG can effectively be placed in contact with it, at the same time increasing the coupling and allowing to use the metallic cavity as one of the gates needed to create the bandgap.\\

\section{Conclusions and perspectives}
\label{Conclusions}
We studied the physics of excitons in GBG resonantly coupled to a photonic microcavity, assessing the possibility to reach the strong and the ultrastrong regimes of cavity quantum electrodynamics, and we discovered that both of them are within experimental reach. We hope that such a result will foster further interest in the study of light-matter coupling in GBG, eventually leading to the realisation of graphene-based polaritonic microcavities. Such devices could empower novel technological breakthroughs in the mid-infrared region of the electromagnetic spectrum, by joining the unique physical properties of tunable gapped graphene and strongly coupled polaritons. 

While the theory presented here is mainly phenomenological, further studies, coadiuvated by ab-initio calculations, could give a clearer picture of the unique properties of those polaritons. In particular it is important to notice that the absolute value of the coupling $\Omega_R$ can become comparable with the excitonic binding energy, thus obliging a consistent theory to consider also the coupling with higher lying excitonic modes and with the continuum of electron states. This kind of effects, previously studied in quantum well-based microcavities \cite{Citrin03,Zhang13,Grenier15}, amounts to a cavity induced modification of the exciton wavefunction, and it leads to an alteration of the spectral features of the polaritonic resonances. 

It is worthwhile to notice that the theory we developed in Sec. \ref{Theory} is based on an effective minimal coupling, inspired to what usually done in the theory of 
microcavity polaritons in quantum wells \cite{Savona94,Kavokin}. Still, a diatribe recently arose on the pertinence of such a model for graphene \cite{Chirolli12,Hagenmuller12}, whose crucial point is the presence, or lack of thereof, of the diamagnetic $\mathbf{A}(\mathbf{r})^2$ term in the minimal coupling Hamiltonian. While the presence of such a term does not change the intensity of the coupling, and it does not thus modify the results of the present work, it still has a non-negligible impact on the system. Not only it shifts the polaritonic energies \cite{Anappara09}, but for high normalised couplings its presence causes an effective decoupling between light and matter \cite{DeLiberato14,Ripoll15}, while its absence leads to an instability of the ground state and a subsequent superradiant phase transition \cite{Lambert04,Nataf10b,Bamba14}. While the solution of this diatribe lies well outside the scope of the present paper, we wish to point out that the very large couplings achievable in GBG polaritons, and the consequent very different predictions for observables with and without the $\mathbf{A}(\mathbf{r})^2$ term, would allow a definitive solution of this conundrum.

\section{Acknowledgements}
The author is Royal Society Research Fellow, and he acknowledges support from EPSRC grant EP/M003183/1.
%The author thanks C.-H. Park, S. Portolan, S. Ribeiro, N. Shammah, and E. Solano for fruitful discussions.

\end{document}